\documentclass[aps,prb,twocolumn,showpacs,amsfonts,a4paper,floatfix]{revtex4-1}

\usepackage{eurosym}
\usepackage{graphicx}% Include figure files

\usepackage[utf8]{inputenc}

\usepackage[sort&compress]{natbib}

\usepackage{bm}          
\usepackage{amsmath}
\usepackage{latexsym}
\usepackage{color}

\graphicspath{{./Figures/}}

\newcommand{\HH}{{\cal H}}

\newcommand{\ket}[1]{{|{#1}\rangle}}
\newcommand{\bra}[1]{{\langle{#1}|}}

\begin{document}

\title{Density matrix renormalization group for a highly degenerate quantum
    system: Sliding environment block approach}

\author{Peter Schmitteckert}
\affiliation{Institute for Theoretical Physics and Astrophysics,
Julius-Maximilian University of W\"urzburg,
Am Hubland, 97074 W\"urzburg,
Germany}
\date{\today}

\begin{abstract}
We present an infinite lattice DMRG sweeping procedure which can be used as a replacement for the standard infinite lattice blocking schemes.
Although the scheme is generally applicable to any system, its main advantages are the correct representation of commensurability issues 
and the treatment of degenerate systems. As an example we apply the method to a spin chain featuring a highly degenerate ground state space
where the new sweeping scheme provides an increase in  performance as well as accuracy  by many orders of magnitude compared to a recently published work.
\end{abstract}

\maketitle

\section{Introduction}
The density matrix renormalization group approach (DMRG) \cite{White:PRL92,White:PRB93,Proceedings98} is one of the most powerful methods 
for low dimensional, actually low entangled, quantum systems. One of its important properties is  that it projects on a subspace of
the complete Hilbert space in which the corresponding linear algebra is performed. It works in a many particle basis and
is therefore perfectly suited to study strongly correlated quantum systems, where the only approximation consists of the size
of the projected subspace, the so-called target space. The DMRG evolved out of Wilson's 
numerical renormalization group scheme  \cite{Wilson:RMP75,Krishnamurthy_Wilkins_Wilson:PRB1980,Krishnamurthy_Wilkins_Wilson:PRB1980b,Bulla_Costi_Pruschke:RMP2008}
(NRG) by realizing that the boundary conditions, and therefore the selection rules, are important for real space blocking schemes \cite{White_Noack:PRL92}.
It turned out \cite{White:PRL92,White:PRB93} that, if we subdivide our system, the so-called superblock $C$, into two parts, blocks $A$ and $B$,
the eigenstates of the reduced density matrices $\rho_{A,B}$ provide a systematic  expansion for the wave function of the system.
Specifically, if an eigenstate $\ket{\Psi}$ of  superblock $C$ is given by
\begin{align}
  \ket{\Psi}  =& \sum_{i,j}  \Psi_{i,j} \,\ket{i}_A \otimes \ket{j}_B \\
        =& \sum_\ell \sigma_\ell  \,\ket{\ell}_{\tilde{A}} \otimes \ket{\ell}_{\tilde{B}} \,,
\end{align}
where $\sigma_\ell^2$ are the eigenvalues of the reduced density matrices 
\begin{align}
  \rho_{A; i,j}  = & \sum_{\ell}  \Psi^*_{i,\ell} \Psi^{}_{j,\ell}  \label{eq:Psi_SVD_A} \\
  \rho_{B;i,j}  = & \sum_{\ell}  \Psi^*_{\ell,i} \Psi^{}_{\ell,j}  \, \label{eq:Psi_SVD_B}
\end{align}
$\Psi_{i,j}$ denotes the wave function with respect to the basis states $\ket{i}_{A(B)}$ of block $A$ ($B$).
From the normalization of $\ket{\Psi}$ it follows that $\sum_\ell \sigma^2_\ell =1$ and since density matrices are semi-positive definite
we have $0 \le \sigma^2_1 \le \sigma^2_2 \le \cdots \le 1$. In addition, $\sigma_\ell$ correspond to the singular values of a singular value decomposition (SVD)
of $\ket{\Psi}$. 
If one keeps the $m$ states with highest $\sigma^2_\ell$, then the discarded entropy 
$
 {\cal S}_{\mathrm d} \,=\, - \sum_{\ell>m} \sigma^2_\ell \,\log \sigma^2_\ell
$
provides a measure of the information that gets projected out. For details see \cite{White:PRL92,White:PRB93,Proceedings98}.
From these observations it is clear, that the eigenstates with the highest eigenvalues of the reduced density matrices are the important
states and that Eqs.~(\ref{eq:Psi_SVD_A},\ref{eq:Psi_SVD_B}) provide a systematic expansion of  $\ket{\Psi}$. 
This key observation combined with a suitable sweeping procedure \cite{White:PRL92,White:PRB93,Proceedings98}
led to the success of the DMRG methods.

Despite this clear foundation of the DMRG it still has the problem of being a Münchhausen (bootstrapping) approach.
Like Münchhausen claimed to be able to pull himself (and his horse!) out of a swamp by pulling at his own hair \cite{Buerger:1786}
the DMRG tries to converge to the true ground state from some initial guess and there is no guarantee that one actually converges towards the ground state.
E.g.\ Ref.~\cite{PS:Proceedings98} provides an example where the DMRG converges to an excited state, provided the number of states per block
is too small, although standard measures, such as the discarded entropy signal perfect convergence. Indeed, there is an excellent convergence to an excited state,
just not to the ground state. This problem is enhanced by the so-called wave function prediction technique \cite{WhitePredicition}
where one seeds the sparse matrix diagonalization of a DMRG step with the results of the preceding step. Whereas this improves the run time significantly,
it also increases the risk of being trapped at an excited state. A way to reduce this risk was provided in Ref.~\cite{PS:Proceedings98}: by adding
the ground state of some homogeneous system to the density matrix during the first sweeps, one can reduce the risk of ending in an excited state significantly,
still there is no guarantee. An alternative idea consisting of adding some mixing terms to the density matrix was suggested in \textcite{White:PRB05}.
In addition, as pointed out in Ref.\cite{Schmitteckert_Werner:PRB04} it is important to include all states $\ket{\Psi_n}$ of
a degenerate ground state, $E_n = E_0$ in the reduced density matrix,
\begin{align}
  \ket{\Psi_n}  =& \sum_{i,j}  \Psi_{n;i,j} \,\ket{i}_A \otimes \ket{j}_B \\
\rho_{A;i,j}  =  \sum_{\ell,n}  \Psi^*_{n;i,\ell} \Psi^{}_{n;j,\ell} & \qquad \rho_{B; i,j}  =  \sum_{\ell,n}  \Psi^*_{n;\ell,i} \Psi^{}_{n;\ell,j}  \,.
\end{align}
Failing to include all states leads to the problem, that the (sparse) matrix diagonalization selects a subspace of the full degenerate ground space only,
which may change in every DMRG step avoiding any convergence.

In the following we describe a method that allows us to efficiently produce high quality 
initial states for the DMRG by accurately keeping the full degenerate subspace.

\section{Sliding block $B$ approach}
\begin{figure}[h]
\includegraphics[width=0.9\columnwidth]{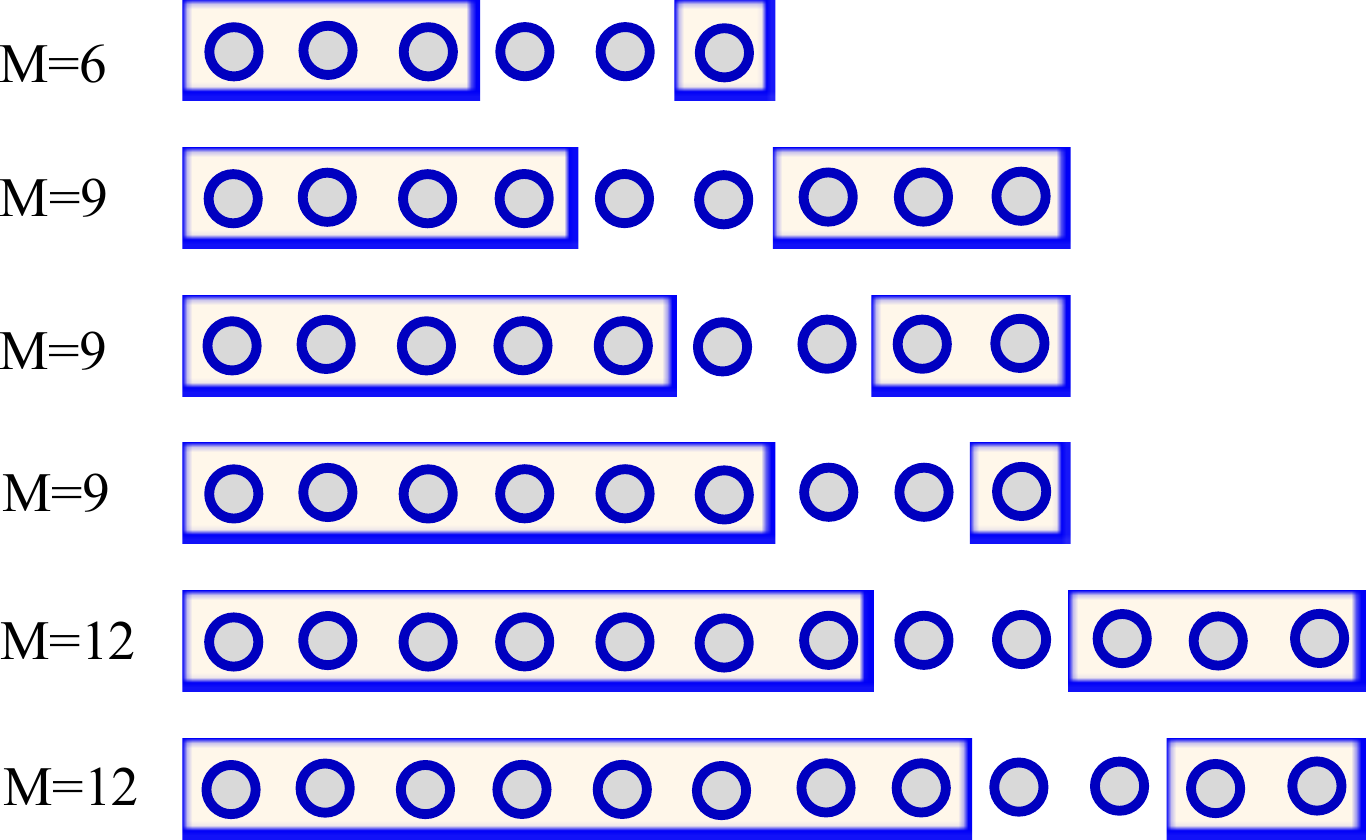}
\caption{The SBB approach to ensure system sizes in multiples of three. Here we start with an initial system of six sites, that gets treated completely.
Next we switch to a $M=9$ site system by increasing the single site environment block  to three sites. We now continue as in a finite lattice
sweeping procedure, until block $B$ consists of one site only. We can then increase block $B$ to three sites again.}
\label{fig:SBB_scheme}
\end{figure}
In a typical DMRG calculation one starts with two blocks $A$ and $B$ that one can still treat exactly,
and builds a superblock by inserting two sites, $A\bullet\bullet B$. One then searches for the ground state of the Hamiltonian in this configuration
and projects on the highest weight states of the reduced density matrices $\rho_{A\bullet}$ and $\rho_{\bullet B}$. By this construction one builds new  
blocks $A$, $B$ where the added site is merged. The dimension of the Hilbert space is now given by
$\dim(A) \dim(\bullet\bullet) \dim(B)$, with $\dim(\bullet\bullet)$ the dimension of the  space of the inserted sites,
and $\dim(A)$ [$\dim(B)$] the dimension of block $A$ [$B$]. Once we started truncation the dimension of block $A$ is given by 
$m$\footnote{Strictly speaking it is at least $m$, as we never cut at degenerate eigenvalues of the reduced density matrices.}. 
The dimension of the target space is the Hilbert space constrained by the explicit quantum numbers.
Note that the diagonalization is performed before truncating block $A\bullet$.

One then continues  increasing the total system size by two sites until the desired system size $M$ is reached. One can now continue
with finite lattice sweeps \cite{White:PRL92,White:PRB93,Proceedings98}, where one keeps the system size fixed,
taking the necessary environment blocks from a previous sweep. Although this approach typically works well,
it is not suited for systems with a commensurate structure that is not given by a period of two sites. For instance, the above procedure
doesn't work well for a charge density wave system with a period larger than two sites, such as a 1D Fermi system with longer ranged interaction \cite{Schmitteckert_Werner:PRB04}.
An alternative scheme for the infinite lattice (warm-up) sweep consists of the sliding block $B$ approach (SBB) 
that was  already successfully applied for fractional quantum hall systems \cite{Hu:PLA12,Johri:PRB14,Johri:NJP16}
and a model of oligo-acenes \cite{PS:JCP2017}. There one also works with a $A\bullet\bullet B$ blocking.
However, only block $A$ is iteratively increased as above. Block $B$ consists of a small number of sites only, which can still be treated completely,
and its size is chosen in order to fulfill commensurability and quantum number constraints. E.g.\ for a 1/3 filled system one can always work with systems sizes which are multiples
of three. In Fig.~\ref{fig:SBB_scheme} we provide a corresponding example. Note, this is only an example. One doesn't have to go down to a single site block $B$,
instead one could also work with block sizes of three, four, and five. The important ingredient is to work with block sizes for $B$ that can always be built from scratch
without the need for referring to earlier iterations as in the standard DMRG infinite lattice sweep procedure.
In the work on oligo--acenes \cite{PS:JCP2017} the SBB warm up was used to ensure that the system always consists of complete unit cells corresponding
to system sizes of $4n+2$ (6, 10, 14, 18, $\cdots$). 
Specifically,  in this case one may choose the block sizes of $(4+2+4)$, $(5+2+3)$, $(6+2+2)$, $(7+2+5)$, $(8+2+4)$, $\cdots$,
where the first number corresponds to the number of sites in block $A$, the 2 for the two inserted sites, and the third number to the number of sites in block $B$.
In this work we want to  demonstrate, that besides commensurability constraints the SBB approach can
also be helpful in the case of highly degenerate systems. As the environment block $B$ can be kept small, e.g.\ consisting of one or two sites only,
the target space can be kept small, simplifying the case of strong degeneracies. 
To this end we study a special case of a spin Hamiltonian and compare to a recent publication by \textcite{Roberts:PRB2017}.

\section{Bravyi-Gosset model}
As an example we look at the Bravyi-Gosset model \cite{BravyiGosset:JMP2015},
\begin{equation}
\HH = \sum_{x=1}^{M-1} \ket{\psi_{x-1,x}} \bra{ \psi_{x-1,x}} \,,
\label{eq:H_BG}
\end{equation}
which consists of a chain of $M$ qubits with hard wall boundary condition (HWBC),
where $\ket{\psi_{x-1,x}}$ is a two qubit state including qubits on sites $x-1$ and $x$. 
Here the Hamiltonian penalizes neighboring qubits to be in the same state,
for details see \textcite{BravyiGosset:JMP2015}. 
For periodic boundary conditions (PBCs) a connection between the first and the last site is added.
The model has the interesting property, that for  a wide range of parameter\cite{BravyiGosset:JMP2015},
the model  possesses an $(M+1)$ degenerate ground state for HWBC, whereas for PBC the ground state space 
can be two or $M+1$ dimensional.
The large degeneracy for HWBC, which may get lifted by the addition of a  single bond may lead to difficulties
with DMRG/matrix product state setups, if the warm-up / infinite lattice sweep is not handled properly.
In the following we show that the SBB protocol solves the problem in an efficient way.

To this end we study the special case of Eq.~\eqref{eq:H_BG} given by a homogeneous chain of maximally entangled qubits\cite{BravyiGosset:JMP2015} in a spin basis,
%%%%%%%%%%%%%%%%%%%%
% Something is broken with the general case, I don't see fow the PRB should be correct.
% \begin{align}
% \HH  =& \sum_{x=1}^{M-1} \hat{S}^z_{x-1} \hat{S}^z_{x} \;+\; \Delta \left( \hat{S}^+_{x-1} \hat{S}^+_{x} \,+\, \hat{S}^-_{x-1} \hat{S}^-_{x} \right) \nonumber \\
%      -& 0.5 \sqrt{ 1 - 4 \Delta^2} \,\sum_{x=0}^{M-1} \hat{S}^z_x  \label{eq:H_XXZ}
% \end{align}
%%%%%%%%%%%%%%%%%
\begin{align}
\HH  =& \sum_{x=1}^{M-1} \hat{S}^z_{x-1} \hat{S}^z_{x} \;+\; \frac{1}{2} \left( \hat{S}^+_{x-1} \hat{S}^+_{x} \,+\, \hat{S}^-_{x-1} \hat{S}^-_{x} \right)  \label{eq:H_XXZ}
\end{align}
with $\hat{S}^\pm_x$, $\hat{S}^z_x$ the standard spin-1/2 ladder and  $z$-component operators at site $x$.
This particular anisotropic Heisenberg model was studied in \textcite{Roberts:PRB2017} for HWBC. 
There it was reported that their DMRG needs
about 40h single core CPU time to obtain the low energy spectrum for $M=32$ sites.
In their results the spectrum is obtained in a successive manner and it is not strictly ordered.
Therefore, they had to consider the 36 lowest eigenstates in order to capture the 33 degenerate ground states.
Their largest deviations from zero of energy differences  in the degenerate ground state space was of the order of $10^{-6}$.
Below we show that DMRG can perform orders of magnitude better, for runtimes as well for system sizes as well as for accuracy.

The Hamiltonian \eqref{eq:H_XXZ} conserves the spin $S_z$ component only modulo 2. That is, for integer spin system, i.e. an even number of sites, 
we only have the quantum numbers $S_z \equiv 0$ and $S_z \equiv 1$, and for half-integer spin sectors, an odd number of sites, 
we have $S_z \equiv 1/2$ and $S_z \equiv 3/2$. 
In fact, we find for even site systems $M/2+1$ states of the degenerate ground state
are in the $S_z \equiv 0$ sector, whereas $M/2$ states are in the $S_z \equiv 1$ sector. For odd system sizes we find
$(M+1)/2$ ground states in each of the two possible sectors. 
Note that in our SBB warm up we either target for odd or for even system sizes. We never change the parity of the system size during the SBB sweep.
In addition we are  exploiting the modulo two spin symmetry and target the corresponding quantum sector in each DMRG step already at each infinite lattice sweep step.

Performing a Jordan-Wigner transformation the model \eqref{eq:H_XXZ} 
maps on to an interacting Kitaev chain \cite{Kitaev2001,ThomaleRachelSchmitteckert2013} without hopping
\begin{align}
\HH  =& \sum_{x=1}^{M-1} \left(\hat{n}_{x-1}-\frac{1}{2}\right) \left( \hat{n}_{x}-\frac{1}{2}\right)  \nonumber\\
     &+\; \frac{1}{2} \sum_{x=1}^{M-1} \left( \hat{c}^\dagger_{x-1} \hat{c}^\dagger_{x} \,+\, \hat{c}^{}_{x-1} \hat{c}^{}_{x} \right)  \label{eq:H_Kitaev}
\end{align}
where $\hat{c}^{}_x$ ($\hat{c}^\dagger_x$) are the standard fermionic annihilation (creation) operators at site $x$, and
$\hat{n}^{}_x = \hat{c}^{\dagger}_x\,\hat{c}^{}_x$ the local density operators.

\section{Hard wall boundary conditions}
\begin{figure}[h]
\includegraphics[width=0.9\columnwidth]{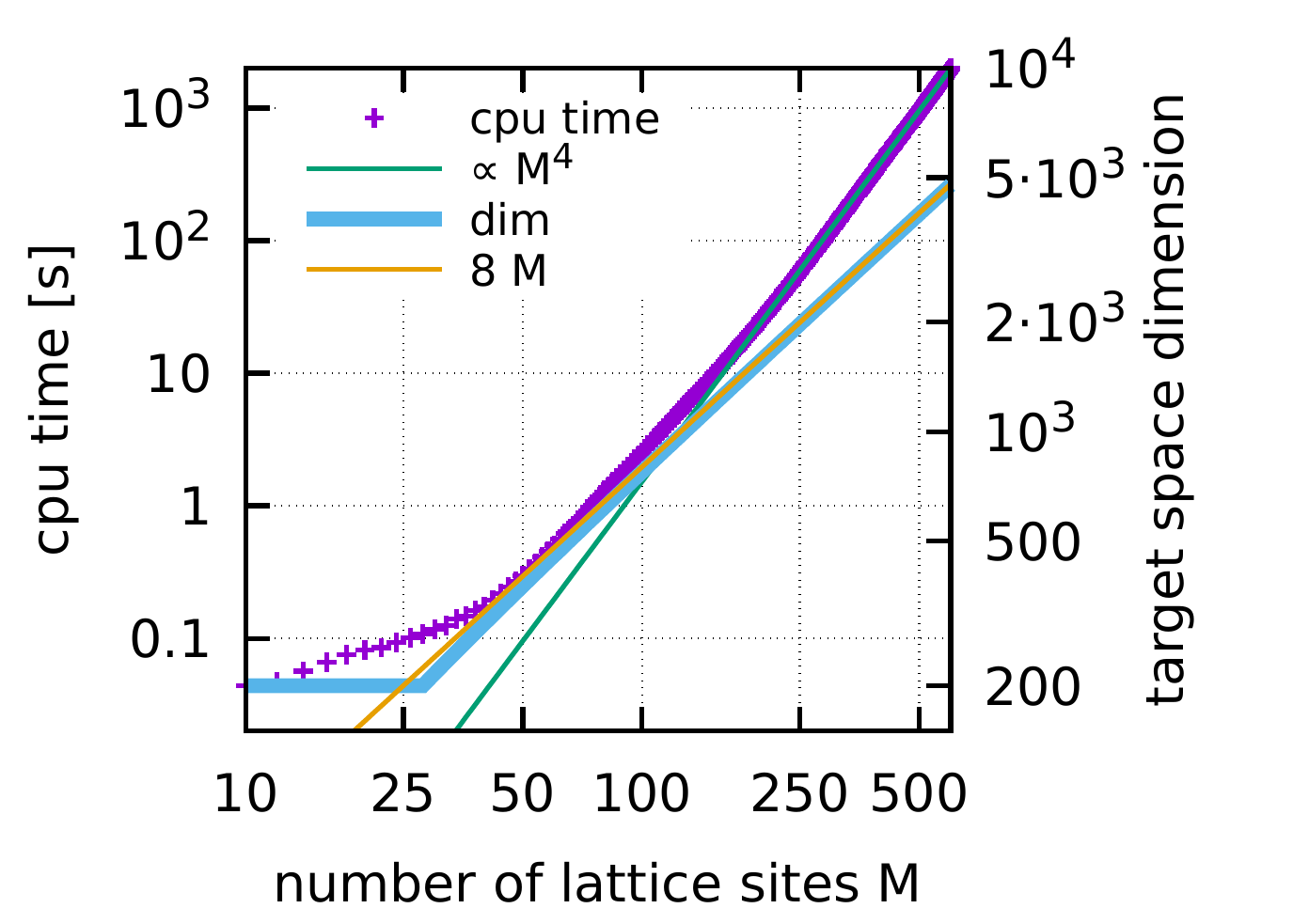}
\caption{Scaling of the CPU time vs.\ the number of lattice sites ranging from $M=10$ to $M=600$ sites, $B_z=0$, for the infinite lattice SBB DMRG
targeting the $M+1$ states lowest in energy for $M$ even. The CPU time is the sum of the CPU time of both spin sectors.
The discarded entropy is enforced below $10^{-12}$,
and at least 25 states are kept in block $A$. Block $B$ is always built exactly. In addition, the target space dimension of the $S_z \equiv0$ is shown, which scales 
as $8 M$, provided the system size $M$ is not too small. Note that  the target space dimensions of the two spin sectors
turn out to be pretty close to each other. 
The highest excitation energy is below $7\cdot10^{-11}$,
and for system sizes below $M=50$ it is below $10^{-13}$. 
}
\label{fig:T}
\end{figure}
First we compare our approach to the results stated in \textcite{Roberts:PRB2017}.
For a comparison we performed a  sliding block $B$ approach as the infinite lattice procedure and seven finite lattice sweeps
tracking the lowest 33 states, keeping enough states per environment block to ensure a discarded entropy below $10^{-12}$ in
each DMRG step. Calculations are performed on a laptop with an Intel E3-1505M CPU and a kernel (Linux 4.15rc7) 
including the kernel page table isolation patches.
Our largest numerical deviation from the true ground state energy $-7.75$ is below $3\cdot 10^{-14}$ and it took less than 27 seconds, 
outperforming \textcite{Roberts:PRB2017} by orders in magnitude for the execution speed as well the accuracy. 
The key to this fast and accurate execution of the code lies in the effectiveness of the infinite lattice sweep.
Indeed, the warm up sweep takes far less than a second, $0.064$ s (0.061 s) in the $S_z\equiv 0$ ($S_z\equiv 1$)  sector,  
and already provides a 33--dimensional subspace with deviations below $5 \cdot 10^{-14}$ from
the true result. That is, the problem is already solved on that level close to machine precision. 
We would like to stress that it is essential to obtain the complete degenerate ground state space  in each DMRG step. 
In failing to keep the complete degenerate ground state space, even by missing just one single state, one spoils the approach.

In Fig.~\ref{fig:T} we show results for the CPU time vs.\ system size for even $M$, i.e.  two sites in block $B$, two insert sites,
and an even number of sites in block $A$. In contrast to \textcite{Roberts:PRB2017} we can easily go to system sizes beyond five hundred sites.
The CPU times presented in Fig.~\ref{fig:T} are the sum of two independent runs for even system sizes,
one for $S_z\equiv 0$, keeping $M/2+1$ low lying states, and one for $S_z \equiv 1$, keeping $M/2$ low lying states.
Even for the 600 site system, the numerical excitation gap for the $601{\text{st}}$ state is below  $7\cdot10^{-11}$,
rendering finite lattice sweeps unnecessary. This could be made even faster, as our code is not optimized for such small target spaces.
We actually build a sparse matrix representation of the Hamiltonian from which we then extract the corresponding dense matrix,
as this is usually only needed at a few initial infinite lattice steps.
In order to calculate correlation functions
one may still want to perform finite lattice sweeps. And indeed, the wave functions of the SBB warm-up are accurate enough
to provide a starting space for an iterative treatment during the finite lattice sweeps, where of course the
target space dimensions grow beyond the applicability of dense matrix methods.
\begin{figure}
\includegraphics[width=0.9\columnwidth]{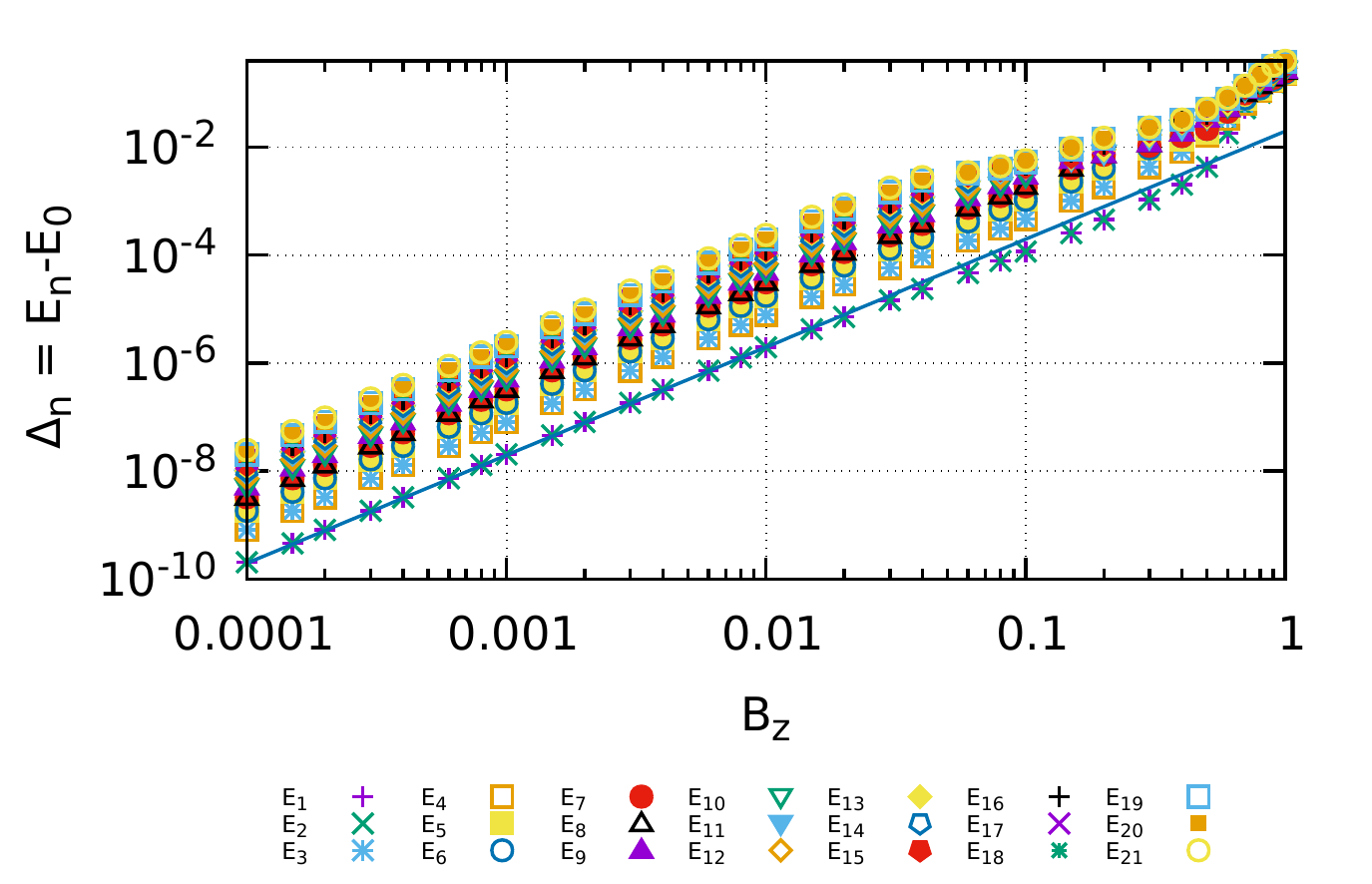}
\caption{Excitation gaps $\Delta_1 \cdots \Delta_{21}$, $\Delta_n = E_n - E_0$, vs.\ magnetic field $B_z$ for a system consisting of $M=50$ sites.
The line is proportional to $B_z^2$. The ground state is in the $S_z \equiv 1$ sector, whereas the first excited state is in the $S_z \equiv 0$ sector.}
\label{fig:En_Bz}
\end{figure}

As said above, in order to achieve those results, it is essential to obtain the complete degenerate subspace at each DMRG step.
However, obtaining hundreds of (nearly--) degenerate states is a non-trivial task and an iterative sparse matrix approach
is hard to get converged. At this point the SBB tremendously simplifies the situation. By using environment blocks
consisting of one or two sites only we can keep the target space small enough in order to apply dense matrix diagonalization
routines, which are stable enough to deal with the degeneracies. In Fig.~\ref{fig:T} we also provide the size of the target space
dimension, which scales only linearly with the system size. This actually points at the true reason for our remarkably fast 
algorithm. The problem appears not to be exponentially hard. At least up to 600 sites, the required target space grows only linearly
with the number of sites. Therefore, the CPU time of each SBB DMRG step grows cubed with the system size resulting in an overall $M^4$ runtime behavior
as observed in Fig.~\ref{fig:T}. The deviation for small system sizes are due to the fact that we kept at least 25 states in block $A$ leading
to a lower bound for the target space dimension.

At finite fields $B_z$, $\HH_B = \HH \,+\, B_z \sum_x \hat{S}^z_x$,
\begin{figure}[t]
\includegraphics[width=0.9\columnwidth]{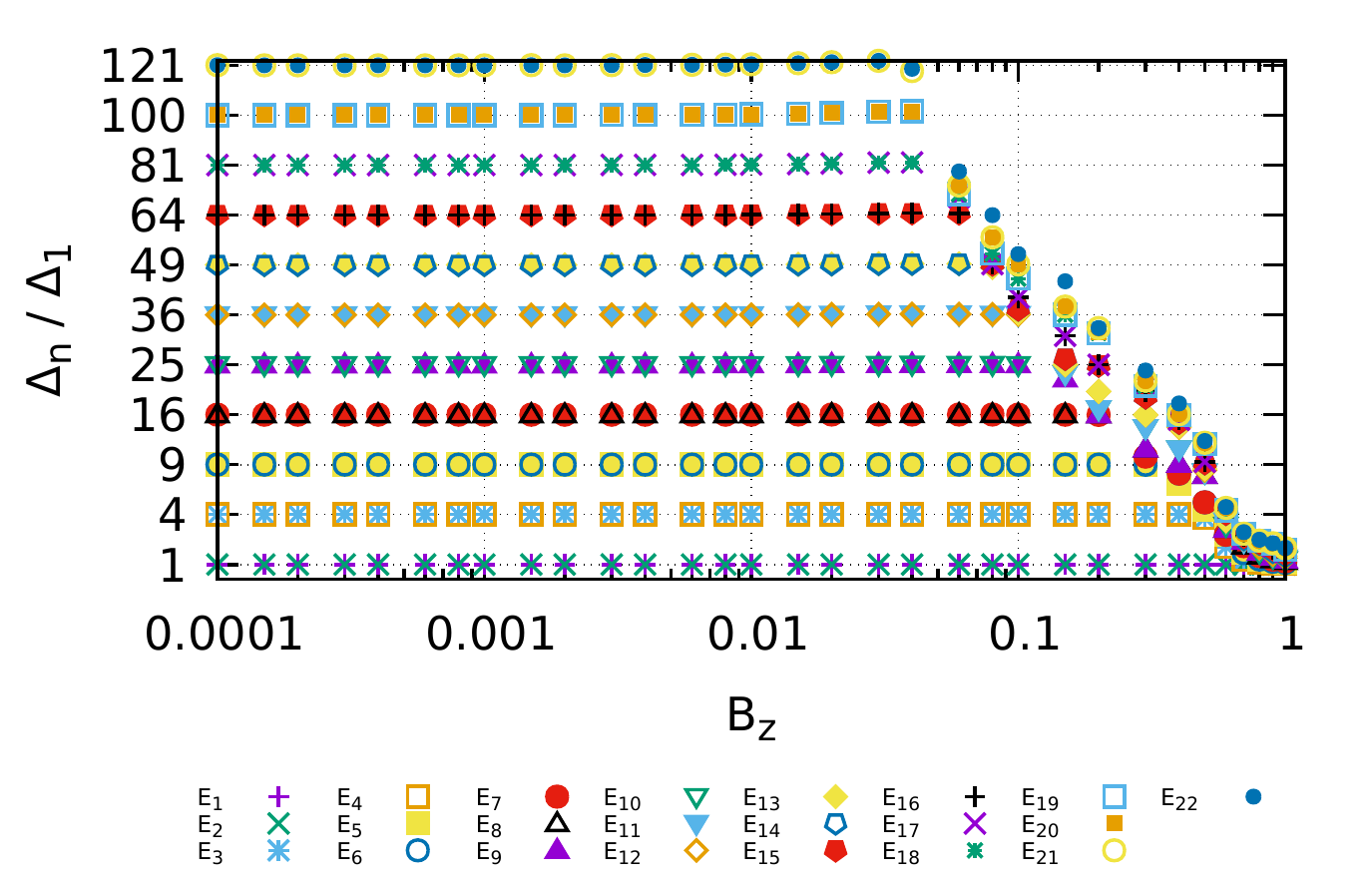}
\caption{Excitation gaps $\Delta_1 \cdots \Delta_{22}$, $\Delta_n = E_n - E_0$, vs. magnetic field $B_z$ for a system consisting of $M=50$ sites
as in Fig.~\ref{fig:En_Bz}. The $y$-axis is scaled by the first excitation gap $\Delta_1$.} %%%=E_1 - E_0$.}
\label{fig:En_scaled}
\end{figure}
we still get very good results from the sliding block $B$ approach.
E.g. the eigenvalue for the $51^\text{st}$ eigenstate of an $M=50$ site system $B_z=0.01$,
is only about $ 0.09\,\%$ higher compared to the full DMRG including finite lattice sweeps. 
It therefore provides an excellent warm up procedure for the DMRG.
In order to illustrate this we show in Fig.~\ref{fig:En_Bz} the low energy spectrum with respect
to an applied magnetic field for a system consisting of $M=50$ sites.
There we performed 7 finite lattice sweeps in addition to the infinite lattice sweep.
Discarded entropy is enforced to be below $10^{-10}$, and the target space dimension grows up to $2.5 \cdot 10^5$
for small $B_z$ fields, and up to  $1.2 \cdot 10^6$ for the larger magnetic fields.
Again the results are obtained by two sets of runs, one for $S_z \equiv 0$ and one for $S_z \equiv 1$.
The results show, that we can obtain a clear quadratic scaling even down to excitation energies below $10^{-9}$.

Finally we show in Fig.~\ref{fig:En_scaled} the excitation gaps for the same system as in Fig.~\ref{fig:En_Bz},
where we normalized the excitation gaps by the first excitation gap. It demonstrates that one can obtain 
a nice quadratic scaling of the excitation gaps for a large set of low lying states.

\section{Periodic boundary conditions}
For PBC and $M$ even, we find the same degeneracy of the ground state as for HWBC:
$M/2+1$ states in  the $S_z \equiv 0$ sector and $M/2$ states in the $S_z \equiv 1$ sector. 
However, for odd system sizes we obtain the remarkable result, which is consistent with \textcite{BravyiGosset:JMP2015}, 
that the ground state is only two-fold degenerate, one state in each of the spin sectors $S_z \equiv 1/2$  and  $S_z \equiv 3/2$.

% \begin{figure}[ht]
% %%\includegraphics[width=1.0\columnwidth]{En_PBC.pdf}
% \caption{Here we show the excitation gaps for PBC.}
% \label{fig:En_PBC}
% \end{figure}

\section{Summary}

In summary we provided a DMRG infinite lattice scheme that provides an improvement of several orders of
magnitude in run-time and accuracy for a specific, highly degenerate spin chain presented in Ref.~\cite{Roberts:PRB2017}.
We pointed out that for those highly degenerate systems it is essential to include the complete degenerate subspace into the density 
matrix to avoid stagnation of the DMRG. In the context of this work the superiority of the SBB stems from the property that
it allows for keeping the target space of the infinite lattice scheme small. Although this appears to be counter--intuitive, it allows
the application of dense matrix methods which can handle large degeneracies in a faithful manner. One can therefore expect that it is
also the preferred scheme in the case of approximately highly degenerate ground state subspaces. 
Finally our results show that the spin chain under investigation does not 
appear to be exponentially hard and it is therefore not surprising that it can be solved in polynomial time.

\section{Acknowledgement}
This work was supported by ERC-StG-Thomale-TOPOLECTRICS-336012.

%%%%%%%%%%%%%%%%%%%%%%%%%%%

\bibliographystyle{apsrev-nourl}

\begin{thebibliography}{22}
\expandafter\ifx\csname natexlab\endcsname\relax\def\natexlab#1{#1}\fi
\expandafter\ifx\csname bibnamefont\endcsname\relax
  \def\bibnamefont#1{#1}\fi
\expandafter\ifx\csname bibfnamefont\endcsname\relax
  \def\bibfnamefont#1{#1}\fi
\expandafter\ifx\csname citenamefont\endcsname\relax
  \def\citenamefont#1{#1}\fi
\expandafter\ifx\csname url\endcsname\relax
  \def\url#1{\texttt{#1}}\fi
\expandafter\ifx\csname urlprefix\endcsname\relax\def\urlprefix{URL }\fi
\providecommand{\bibinfo}[2]{#2}
\providecommand{\eprint}[2][]{\url{#2}}

\bibitem[{\citenamefont{White}(1992)}]{White:PRL92}
\bibinfo{author}{\bibfnamefont{S.~R.} \bibnamefont{White}},
  \bibinfo{journal}{Phys. Rev. Lett.} \textbf{\bibinfo{volume}{69}},
  \bibinfo{pages}{2863} (\bibinfo{year}{1992}).

\bibitem[{\citenamefont{White}(1993)}]{White:PRB93}
\bibinfo{author}{\bibfnamefont{S.~R.} \bibnamefont{White}},
  \bibinfo{journal}{Phys.\ Rev.\ B} \textbf{\bibinfo{volume}{48}},
  \bibinfo{pages}{10345} (\bibinfo{year}{1993}).

\bibitem[{\citenamefont{Peschel et~al.}(1999)\citenamefont{Peschel, Wang,
  M.Kaulke, and Hallberg}}]{Proceedings98}
\bibinfo{editor}{\bibfnamefont{I.}~\bibnamefont{Peschel}},
  \bibinfo{editor}{\bibfnamefont{X.}~\bibnamefont{Wang}},
  \bibinfo{editor}{\bibnamefont{M.Kaulke}}, \bibnamefont{and}
  \bibinfo{editor}{\bibfnamefont{K.}~\bibnamefont{Hallberg}}, eds.,
  \emph{\bibinfo{title}{Density Matrix Renormalization}}
  (\bibinfo{year}{1999}), ISBN \bibinfo{isbn}{978-3-540-66129-0}.

\bibitem[{\citenamefont{Wilson}(1975)}]{Wilson:RMP75}
\bibinfo{author}{\bibfnamefont{K.~G.} \bibnamefont{Wilson}},
  \bibinfo{journal}{Rev. Mod. Phys.} \textbf{\bibinfo{volume}{47}},
  \bibinfo{pages}{773} (\bibinfo{year}{1975}).

\bibitem[{\citenamefont{Krishna-murthy
  et~al.}(1980{\natexlab{a}})\citenamefont{Krishna-murthy, Wilkins, and
  Wilson}}]{Krishnamurthy_Wilkins_Wilson:PRB1980}
\bibinfo{author}{\bibfnamefont{H.~R.} \bibnamefont{Krishna-murthy}},
  \bibinfo{author}{\bibfnamefont{J.~W.} \bibnamefont{Wilkins}},
  \bibnamefont{and} \bibinfo{author}{\bibfnamefont{K.~G.}
  \bibnamefont{Wilson}}, \bibinfo{journal}{Phys. Rev. B}
  \textbf{\bibinfo{volume}{21}}, \bibinfo{pages}{1003}
  (\bibinfo{year}{1980}{\natexlab{a}}).

\bibitem[{\citenamefont{Krishna-murthy
  et~al.}(1980{\natexlab{b}})\citenamefont{Krishna-murthy, Wilkins, and
  Wilson}}]{Krishnamurthy_Wilkins_Wilson:PRB1980b}
\bibinfo{author}{\bibfnamefont{H.~R.} \bibnamefont{Krishna-murthy}},
  \bibinfo{author}{\bibfnamefont{J.~W.} \bibnamefont{Wilkins}},
  \bibnamefont{and} \bibinfo{author}{\bibfnamefont{K.~G.}
  \bibnamefont{Wilson}}, \bibinfo{journal}{Phys. Rev. B}
  \textbf{\bibinfo{volume}{21}}, \bibinfo{pages}{1044}
  (\bibinfo{year}{1980}{\natexlab{b}}).

\bibitem[{\citenamefont{Bulla et~al.}(2008)\citenamefont{Bulla, Costi, and
  Pruschke}}]{Bulla_Costi_Pruschke:RMP2008}
\bibinfo{author}{\bibfnamefont{R.}~\bibnamefont{Bulla}},
  \bibinfo{author}{\bibfnamefont{T.~A.} \bibnamefont{Costi}}, \bibnamefont{and}
  \bibinfo{author}{\bibfnamefont{T.}~\bibnamefont{Pruschke}},
  \bibinfo{journal}{Rev. Mod. Phys.} \textbf{\bibinfo{volume}{80}},
  \bibinfo{pages}{395} (\bibinfo{year}{2008}).

\bibitem[{\citenamefont{White and Noack}(1992)}]{White_Noack:PRL92}
\bibinfo{author}{\bibfnamefont{S.~R.} \bibnamefont{White}} \bibnamefont{and}
  \bibinfo{author}{\bibfnamefont{R.~M.} \bibnamefont{Noack}},
  \bibinfo{journal}{Phys. Rev. Lett.} \textbf{\bibinfo{volume}{68}},
  \bibinfo{pages}{3487} (\bibinfo{year}{1992}).

\bibitem[{\citenamefont{B{\"u}rger}(1786)}]{Buerger:1786}
\bibinfo{author}{\bibfnamefont{G.~A.} \bibnamefont{B{\"u}rger}},
  \emph{\bibinfo{title}{Des Freyherrn von M{\"u}nchhausen Wunderbare Reisen}}
  (\bibinfo{publisher}{Johann Christian Dieterich}, \bibinfo{address}{London
  [G{\"o}ttingen], https://de.wikisource.org}, \bibinfo{year}{1786}).

\bibitem[{\citenamefont{Schmitteckert}(1999)}]{PS:Proceedings98}
\bibinfo{author}{\bibfnamefont{P.}~\bibnamefont{Schmitteckert}}, in
  \emph{\bibinfo{booktitle}{Density Matrix
  Renormalization\cite{Proceedings98}}} (\bibinfo{year}{1999}), pp.
  \bibinfo{pages}{345--355}, ISBN \bibinfo{isbn}{978-3-540-66129-0}.

\bibitem[{\citenamefont{White}(1996)}]{WhitePredicition}
\bibinfo{author}{\bibfnamefont{S.~R.} \bibnamefont{White}},
  \bibinfo{journal}{Phys.\ Rev.\ Lett} \textbf{\bibinfo{volume}{77}},
  \bibinfo{pages}{3633} (\bibinfo{year}{1996}).

\bibitem[{\citenamefont{White}(2005)}]{White:PRB05}
\bibinfo{author}{\bibfnamefont{S.~R.} \bibnamefont{White}},
  \bibinfo{journal}{Phys. Rev. B} \textbf{\bibinfo{volume}{72}},
  \bibinfo{pages}{180403(R)} (\bibinfo{year}{2005}).

\bibitem[{\citenamefont{Schmitteckert and
  Werner}(2004)}]{Schmitteckert_Werner:PRB04}
\bibinfo{author}{\bibfnamefont{P.}~\bibnamefont{Schmitteckert}}
  \bibnamefont{and} \bibinfo{author}{\bibfnamefont{R.}~\bibnamefont{Werner}},
  \bibinfo{journal}{Phys. Rev. B} \textbf{\bibinfo{volume}{69}},
  \bibinfo{pages}{195115} (\bibinfo{year}{2004}).

\bibitem[{Note1()}]{Note1}
Note1, \bibinfo{note}{strictly speaking it is at least $m$, as we never cut at
  degenerate eigenvalues of the reduced density matrices.}

\bibitem[{\citenamefont{Hu et~al.}(2012)\citenamefont{Hu, Papic, Johri, Bhatt,
  and Schmitteckert}}]{Hu:PLA12}
\bibinfo{author}{\bibfnamefont{Z.-X.} \bibnamefont{Hu}},
  \bibinfo{author}{\bibfnamefont{Z.}~\bibnamefont{Papic}},
  \bibinfo{author}{\bibfnamefont{S.}~\bibnamefont{Johri}},
  \bibinfo{author}{\bibfnamefont{R.~N.} \bibnamefont{Bhatt}}, \bibnamefont{and}
  \bibinfo{author}{\bibfnamefont{P.}~\bibnamefont{Schmitteckert}},
  \bibinfo{journal}{Phys. Lett. A} \textbf{\bibinfo{volume}{376}},
  \bibinfo{pages}{2157} (\bibinfo{year}{2012}).

\bibitem[{\citenamefont{Johri et~al.}(2014)\citenamefont{Johri, Papic, Bhatt,
  and Schmitteckert}}]{Johri:PRB14}
\bibinfo{author}{\bibfnamefont{S.}~\bibnamefont{Johri}},
  \bibinfo{author}{\bibfnamefont{Z.}~\bibnamefont{Papic}},
  \bibinfo{author}{\bibfnamefont{R.~N.} \bibnamefont{Bhatt}}, \bibnamefont{and}
  \bibinfo{author}{\bibfnamefont{P.}~\bibnamefont{Schmitteckert}},
  \bibinfo{journal}{Phys. Rev. B} \textbf{\bibinfo{volume}{89}},
  \bibinfo{pages}{115124} (\bibinfo{year}{2014}).

\bibitem[{\citenamefont{Johri et~al.}(2016)\citenamefont{Johri, Papic,
  Schmitteckert, Bhatt, and Haldane}}]{Johri:NJP16}
\bibinfo{author}{\bibfnamefont{S.}~\bibnamefont{Johri}},
  \bibinfo{author}{\bibfnamefont{Z.}~\bibnamefont{Papic}},
  \bibinfo{author}{\bibfnamefont{P.}~\bibnamefont{Schmitteckert}},
  \bibinfo{author}{\bibfnamefont{R.~N.} \bibnamefont{Bhatt}}, \bibnamefont{and}
  \bibinfo{author}{\bibfnamefont{F.~D.~M.} \bibnamefont{Haldane}},
  \bibinfo{journal}{NJP} \textbf{\bibinfo{volume}{18}}, \bibinfo{pages}{025011}
  (\bibinfo{year}{2016}).

\bibitem[{\citenamefont{Schmitteckert et~al.}(2017)\citenamefont{Schmitteckert,
  Thomale, Koryt{\'a}r, and Evers}}]{PS:JCP2017}
\bibinfo{author}{\bibfnamefont{P.}~\bibnamefont{Schmitteckert}},
  \bibinfo{author}{\bibfnamefont{R.}~\bibnamefont{Thomale}},
  \bibinfo{author}{\bibfnamefont{R.}~\bibnamefont{Koryt{\'a}r}},
  \bibnamefont{and} \bibinfo{author}{\bibfnamefont{F.}~\bibnamefont{Evers}},
  \bibinfo{journal}{The Journal of Chemical Physics}
  \textbf{\bibinfo{volume}{146}}, \bibinfo{pages}{092320}
  (\bibinfo{year}{2017}).

\bibitem[{\citenamefont{Roberts et~al.}(2017)\citenamefont{Roberts, Vidick, and
  Motrunich}}]{Roberts:PRB2017}
\bibinfo{author}{\bibfnamefont{B.}~\bibnamefont{Roberts}},
  \bibinfo{author}{\bibfnamefont{T.}~\bibnamefont{Vidick}}, \bibnamefont{and}
  \bibinfo{author}{\bibfnamefont{O.~I.} \bibnamefont{Motrunich}},
  \bibinfo{journal}{Phys. Rev. B} \textbf{\bibinfo{volume}{96}},
  \bibinfo{pages}{214203} (\bibinfo{year}{2017}).

\bibitem[{\citenamefont{Bravyi and Gosset}(2015)}]{BravyiGosset:JMP2015}
\bibinfo{author}{\bibfnamefont{S.}~\bibnamefont{Bravyi}} \bibnamefont{and}
  \bibinfo{author}{\bibfnamefont{D.}~\bibnamefont{Gosset}},
  \bibinfo{journal}{Journal of Mathematical Physics}
  \textbf{\bibinfo{volume}{56}}, \bibinfo{pages}{061902}
  (\bibinfo{year}{2015}).

\bibitem[{\citenamefont{Kitaev}(2001)}]{Kitaev2001}
\bibinfo{author}{\bibfnamefont{A.~Y.} \bibnamefont{Kitaev}},
  \bibinfo{journal}{Physics-Uspekhi} \textbf{\bibinfo{volume}{44}},
  \bibinfo{pages}{131} (\bibinfo{year}{2001}).

\bibitem[{\citenamefont{Thomale et~al.}(2013)\citenamefont{Thomale, Rachel, and
  Schmitteckert}}]{ThomaleRachelSchmitteckert2013}
\bibinfo{author}{\bibfnamefont{R.}~\bibnamefont{Thomale}},
  \bibinfo{author}{\bibfnamefont{S.}~\bibnamefont{Rachel}}, \bibnamefont{and}
  \bibinfo{author}{\bibfnamefont{P.}~\bibnamefont{Schmitteckert}},
  \bibinfo{journal}{Phys. Rev. B} \textbf{\bibinfo{volume}{88}},
  \bibinfo{pages}{161103} (\bibinfo{year}{2013}).

\end{thebibliography}

\end{document}